\documentclass[11pt,tightenlines,eqsecnum,floats,aps,amsmath,amssymb,nofootinbib,prd,shownopacs,notitlepage,longbibliography]{revtex4-2}

\usepackage{ragged2e}
\usepackage{graphicx}
\usepackage[utf8]{inputenc}
\usepackage{amsmath,amsfonts,amssymb}
\usepackage{color}
\usepackage[hidelinks]{hyperref}
\hypersetup{
  colorlinks   = true,
  urlcolor     = blue, 
  linkcolor    = blue,
  citecolor   = blue 
}

\def\Mpc{{\rm Mpc}}

\def \mpl{{\rm m}_{\rm Pl}}
\def \tpl{{\rm t}_{\rm Pl}}

\begin{document}

\title{Primordial magnetogenesis in loop quantum cosmology}
\author{Ganga R. Nair}
\email{ganga.227ph002@nitk.edu.in}
\author{V. Sreenath}
\email{sreenath@nitk.edu.in}
\affiliation{Department of Physics, National Institute of Technology Karnataka, Surathkal, Mangaluru 575025, India.}

\begin{abstract}
 Primordial magnetic fields (PMFs) are magnetic fields generated during the early Universe. These fields are thought to be the seeds of extragalactic magnetic fields. The origin of PMFs is not well known. Further, if they are indeed sources of extragalactic fields, then there is a possibility that observations of extragalactic magnetic fields could provide insights into the primordial physics. With this motivation, we study the generation of the primordial magnetic field in the context of loop quantum cosmology (LQC). In LQC, inflation is preceded by a quantum bounce. 
 In this work, we consider an electromagnetic field coupled to the background as a test field and study its evolution through the bounce and through the subsequent inflationary phase.
 We investigate the power spectra generated in LQC and show that it is scale-dependent.
 We study the power spectra with different initial conditions, discuss equivalent forms of coupling functions, investigate backreaction, and compute the amount of primordial magnetic field which can be measured today. We conclude the article with a summary and discussion of the results. 
\end{abstract}

\maketitle

\section{Introduction}
Our Universe is governed by four fundamental forces. Of these gravitational and electromagnetic forces can exist over long distances. Gravitation, described by the general theory of relativity, forms the basis of the standard model of cosmology. It is interesting to investigate the presence of long-range electromagnetic fields and the role they play in the evolution of our Universe. Charge neutrality of our Universe implies that there are no long-range electric fields. However, the presence of magnetic fields has been observed at all scales. Magnetic fields are known to exist in Earth \cite{OLSON20149} and other planets \cite{Stevenson_planetaryMF}, the Sun \cite{hathaway2015solar, vasilsolardynamo} and other stars \cite{DonatiMFstars}, galaxies \cite{Kronberg_1994_EGMF,beck2001galactic}, clusters of galaxies \cite{Widrow_Lawrence, ClusterMF} and even in cosmic voids \cite{neronov2010evidence}. Magnetic fields present in various astrophysical objects vary in strength and extent. Magnetic fields of the order of a few micro gauss have been measured in spiral galaxies, coherent on scales up to ten kiloparsec \cite{Beck2013Wielebinski}. Similar magnetic fields are also known to be present in clusters of galaxies \cite{MFinGClusters:strength}. The presence of magnetic fields of the order of $10^{-16}$ G  has been observed in cosmic voids \cite{TaylorEGMFblazar}. While the existence of magnetic fields in astrophysical objects can be attributed to mechanisms such as battery and dynamo \cite{Durrer:2013pga, Brandenburg_Subramanyan2005}, their presence in voids is intriguing. 
Observations of TeV blazars by Fermi/LAT and HESS suggest the presence of magnetic fields of order of $10^{-15}$ G in the intergalactic medium \cite{Tavecchio:2010mk}. 

\par 
It has been shown that electromagnetic fields of strengths observed in intergalactic voids can be generated in the early Universe during inflation \cite{Turner_Widrow_1988, Ratra:1991bn,  Lemoine_String95, Bamba:dilatonEM, Campanelli:2007cg, Martin_2008, Subramanian:2009, Watanabe:2010fh, RuthDurrer_2011, Gasperni_graviphotons, Agullo:2013tba, RSharma:helical2018, Sagarika_2022,  Kushwaha_2023, Pal:2023DebMaity} or other primordial scenarios \cite{Hogan1983, Dolgov:Silk1993, Baym:1995EWPT, Brandenburg:1996fc, Dolgov:Grasso2001, Sriramkumar:2015yza, Chowdhury:2016aet} (for texts and reviews, see, for instance, \cite{1979cmft.book.....P, 2021amff.book.....S, Widrow_2011, Durrer:2013pga, Subramanian:2015lua, Vachaspati:2020}). See also \cite{Garg:2025mcc, Ghosh:2025bqp} in this context. In a spatially flat Friedmann-Lema\^itre-Robertson-Walker (FLRW) universe, standard electromagnetic action is conformally invariant. Hence, the electromagnetic fields are not affected by the spacetime curvature of the universe. This implies that as the universe expands nearly exponentially during inflation, the energy density of magnetic fields will get diluted very quickly. This difficulty in generating magnetic fields can be overcome if we break the conformal invariance. There are various ways to break conformal invariance; one simple way is to couple the electromagnetic field to the background dynamics of the scalar field that drives the primordial universe, namely the inflaton (see, for instance, \cite{Turner_Widrow_1988, Ratra:1991bn, Martin_2008,  Subramanian:2009}). This will ensure that magnetic fields present in the early Universe will not get diluted during inflation. In this scenario, magnetic fields provide a new window to learn about the early Universe through their coupling to the background. 
\par 
Loop quantum cosmology (LQC) is an attempt to extend the inflationary phase to the Planck regime \cite{Bojowald:2001xe, PhysRevLett.96.141301, Ashtekar:2006uz, Ashtekar:2006wn, Ashtekar:2009mm, Ashtekar:2007em, Agullo:2012sh, Agullo:2012fc, Agullo:2013ai, Diener:2013uka, Diener:2014mia, Agullo:2015tca, PhysRevD.97.066021, Sreenath:2019uuo, K:2023gsi, K:2024sla}. For reviews on LQC, refer, for instance, \cite{Ashtekar:2011ni, Agullo:2023rqq}. In LQC, inflation is preceded by a quantum bounce. The study of the magnetic field in LQC is interesting due to at least two reasons. First, it would be interesting to understand whether the magnetic field generated in LQC can lead to the observed magnetic fields. Second, magnetic fields generated in LQC could carry signatures of the quantum bounce. These signatures could provide vital information about the Planck regime. 
If tiny anisotropies are present in the background, they could grow during the contracting phase, leading to the bounce, and the spacetime during the quantum bounce could be anisotropic (see, for instance, \cite{Ashtekar:2009vc, Gupt:2012vi, Gupt:2013swa, Agullo:2020kil, Agullo:2020uii, Agullo:2020iqv, Agullo:2020wur}). There have been efforts to understand the generation of homogeneous magnetic fields in this context (see, for instance, \cite{Maartens:2008dd, Rikhvitsky:2013pfa, Motaharfar:2024afn}). In this article, we shall take a different approach and study the generation of an inhomogeneous magnetic field in LQC, assuming a homogeneous and isotropic spacetime. We shall consider the electromagnetic field as a test field coupled to the background dynamics in LQC. 
\par 
This article is organised as follows. In section \ref{sec:2}, we briefly review the treatment of a magnetic field in a spatially flat FLRW spacetime. In section \ref{sec:3}, we discuss the essential aspects of LQC and the details of the potentials of the background scalar field that we consider. In section \ref{sec:4}, we discuss the evolution of magnetic fields in LQC and present our numerical calculations of the same. 
We then analyse different aspects of magnetogenesis, including the effect of changing the time at which initial conditions are imposed, different forms of coupling functions, backreaction and the amount of primordial magnetic field that can be observed today in section \ref{sec:5}. We conclude this article with a summary and discussion of results in section \ref{sec:6}.
\par 
We shall work with natural units, {\it i.e.}, $c\,=\,\hbar\, =\, 1$. All masses will be expressed in Planck mass $\mpl\,=\, 1/\sqrt{G}$, where $G$ is Newton's gravitational constant. 
Depending upon convenience, we shall work with cosmic time ($t$), conformal time ($\eta$), or e-fold ($N$). Different times are related to each other through ${\rm d}N\, =\,H(t)\,{\rm d}t\, =\, a(\eta)\,H(\eta)\,{\rm d}\eta$, where $a$ is the scale factor and $H$ is the Hubble parameter. For brevity, whenever there is no ambiguity, we shall omit explicitly expressing the time or spatial dependence. An overdot and prime will refer to a derivative with respect to cosmic time and conformal time, respectively. Throughout this article, we will use the suffixes `e and 0' to denote a quantity computed at the end of inflation and today, respectively. For instance, $a_e$ and $a_0$ will refer to the value of the scale factor at the end of inflation and today, respectively. Greek indices such as $\mu,\,\nu,\dots$ imply either a time or spatial index. Latin indices such as $i,\,j,\,\dots$ will indicate only the spatial index. Finally, the magnetic field will be expressed in units of gauss (G).

\section{Magnetogenesis in FLRW spacetime \label{sec:2}}
Consider a spatially flat FLRW spacetime. We shall assume that the background dynamics is sourced by a scalar field $\phi(t)$. Let $A_\mu (t, \vec x)$ be the 4-vector potential that describes a test electromagnetic field living on this background. Such an electromagnetic field can be described by the action \cite{Turner_Widrow_1988, Ratra:1991bn, Martin_2008, Subramanian:2009} 
\begin{equation}\label{eqn:action}
    S = - \frac{1}{4} \int {\rm d}t\,\int\, {\rm d}^3x\, a^3\, g^{\alpha \beta} g^{\mu \nu} f^2(t) F_{\mu \alpha} F_{\nu \beta}\, ,
\end{equation}
where $F_{\mu\nu}\,=\, \partial_\mu A_\nu\,-\,\partial_\nu\,A_\mu$ is the electromagnetic field tensor and $f(t)$ is a background-dependent function which couples the electromagnetic field to the background fields. 
\par 
The equation of motion of the vector potential is
\begin{equation}
    \partial_\mu\,\biggl[\,a^3\,f^2(t)\,F^{\mu\nu} \biggr]\,=\,0.
\end{equation}
In Coulomb gauge, {\it i.e.}, $\partial_\mu A^\mu(t, \vec x)\,=0$ and $A_0(t,\,\vec x)\,=0$, the equation of motion becomes
\begin{equation}\label{eqn:Ai}
    \ddot A_i\, +\, \left( H\,+\, \frac{2\dot f}{f}\right) \dot A_i\,-\,\partial^2 A_i\,=\,0,
\end{equation}
where $\partial^2\,=\,a^{-2}\,\delta^{ij}\partial_i\partial_j$. 
The electric and magnetic fields are related to the vector potential, in Coulomb gauge, as $E_i\, =\, -\dot A_i$ and $B_i\,=\, \eta_{ijk}\,\partial_jA_k/a$ respectively, where $\eta_{ijk}$ is the completely antisymmetric tensor with $\eta_{123}\,=\,1$. 
\par 
If we consider the 4-vector potential as a quantum field, vacuum fluctuations occur. The field thus created through quantum fluctuations, if suitably coupled to the background, will grow and possibly lead to the magnetic field pervading our universe. In Coulomb gauge, the vector potential $A_i$ can be quantized by promoting it and its momentum $\Pi_i(t,\vec x)\,=\,a^3(t)\,f^2(t)\,\dot A_i(t,\vec x)$ to operators and the canonical Poisson bracket to quantum commutator. The canonical Poisson bracket is given by 
\begin{equation}
\left\{A^i(t,\vec x),\, \Pi_j(t,\vec y)\right\}\, =\, 
i\,\int \frac{{\rm d}^3\,\vec k}{(2\,\pi)^{3/2}}\, e^{i\,\vec k \cdot (\vec x\,-\, \vec y)} \left( \delta^i_j\, -\, \delta_{jl}\frac{k^i\,k^l}{k^2} \right),
\end{equation}
where $k$ is the comoving wave number. Though vector potential is a 4-vector, it has only two degrees of freedom. To extract the physical degrees of freedom, let us express the vector potential in the following orthonormal basis: $\epsilon^\mu_0\,=\, (1/a,\, 0,\, 0,\, 0)$, $\epsilon^\mu_\lambda\,=\, (0, \tilde\epsilon^i_\lambda/a)$ with $\lambda\,\in \,[1,\,2]$ and $\epsilon^\mu_3\,=\,(0,k^i/(a\,k))$. Here, $\delta_{ij}\,\tilde\epsilon^i_\lambda\,\tilde\epsilon^j_\lambda\,=\,1$. If we expand $A_i(t,\,\vec x)$ in terms of these basis vectors, in Coulomb gauge, only components along $\epsilon^\mu_\lambda$ will exist. If $A(t,\, k)$ is the Fourier mode satisfying the equation of motion of the vector potential, then the quantum operator 
\begin{equation}
    \hat A_i(t,\, k)\, =\, \int \frac{{\rm d}^3\vec k}{(2\,\pi)^{3/2}}\, \sum_{\lambda\,=\,1}^2\, \epsilon_{i\,\lambda}\,(\vec k)\, \left[\,\hat  b_\lambda (k)\,A(t,\vec k)\, e^{i\,{\vec k} \cdot {\vec x}}\, +\,\hat  b^\dagger_\lambda (k)\,A^*(t,\vec k)\, e^{-i\,{\vec k} \cdot {\vec x}} \right].
\end{equation}
The operators $\hat b_\lambda(\vec k)$ and $\hat b^\dagger_\lambda(\vec k)$ are annihilation and creation operators. They satisfy the commutation relation $\left[\,\hat b_\lambda(\vec k), \hat b^\dagger_\lambda(\vec k) \right]\,=\, 1$, with all other possible commutations being zero.
\par 
The expectation value of the energy density of the electric and magnetic fields generated in this process can be found from the respective parts of the `$00$' component of the energy-momentum tensor. The energy density of the magnetic field can be computed to be 
\begin{equation}\label{eqn:rhoB}
\rho^B(t)\,=\, -\langle 0|\,T^{B0}_{0} |0\rangle\, =\, \frac{f^2(t)}{a^2}\int \frac{{\rm d}^3\vec k}{(2\pi)^3}\, k^2\, |A(t, {\vec k})|^2, 
\end{equation}
where $|0\rangle$ is the vacuum annihilated by $b_\lambda (k)$ for all $\lambda$ and $\vec k$. The amount of generated magnetic field can also be quantified in terms of power spectrum, which is the energy density of the magnetic field per logarithmic interval of wave number. It is given by
\begin{equation} \label{eqn:PB}
    {\cal P}^B(t,\,k)\, =\, \frac{{\rm d}\rho^B(t)}{{\rm d}\ln k}\,=\, \frac{k^5}{2\pi^2}\, \left\vert\frac{f(t)\,A(t,k)}{a}\right\vert^2.
\end{equation}
Similarly, one can compute the expectation value of the energy density of the electric field to be 
\begin{equation} \label{eqn:rhoE}
    \rho^E(t)\,=\,-\langle 0|\,T^{E0}_0 \,|0\rangle\,=\, f^2(t)\,\int \frac{{\rm d}^3\vec k}{(2\pi)^3}\, |H\,A(t,\,\vec{k})\,+\,\dot A(t, {\vec k})|^2.
\end{equation}
The power spectrum of the electric field is
\begin{equation} \label{eqn:PE}
    {\cal P}^E(t,\,k)\, =\, \frac{{\rm d}\rho^E(t)}{{\rm d}\ln k}\,=\, \frac{k^3}{2\,\pi^2}\, f^2(t)\,\left|H\,A(t,\,\vec{k})\,+\,\dot A(t, {\vec k})\,\right|^2.
\end{equation}
\section{Loop Quantum Cosmology \label{sec:3}}
Loop quantum cosmology is an attempt, based on the principles of loop quantum gravity, to extend the inflationary scenario into the Planck regime (for reviews, see, for instance, \cite{Ashtekar:2011ni, Agullo:2023rqq}). 
In LQC, a FLRW spacetime sourced by a scalar field is defined by a quantum wave function, which depends only on the scalar field and the volume, where volume is proportional to the cube of the scale factor. We will consider the case where energy density in the quantum regime is dominated by the kinetic term of the scalar field \cite{PhysRevLett.96.141301, Ashtekar:2007em, Ashtekar:2011ni}. As we will see, such kinetic-dominated bounces could lead to sufficient e-folds of inflation after the bounce \cite{Ashtekar:2011rm, Bolliet:2017czc, Bhardwaj:2018omt}. 
Among the class of solutions admitted by LQC, we are interested in states that are sharply peaked around the classical trajectories, particularly those with small dispersion in volume, at late times. It has been shown that \cite{Ashtekar:2006uz, Ashtekar:2006wn, Diener:2013uka, Diener:2014mia, Agullo:2016hap}, for such states, the quantum dynamics are effectively captured by the modified Friedmann-Raychaudhuri equations, {\it viz.}, 
\begin{subequations}\label{eqn:lqc}
\begin{align}
    \left( \frac{\dot a}{a} \right)^2 &= \frac{8 \pi}{3\,\mpl^2} \rho \left( 1 - \frac{\rho}{\rho_{\rm sup}}\right),\\
    \frac{\ddot a}{a}\, &= - \frac{4 \pi }{3\,\mpl^2} \rho \left(1 - 4 \frac{\rho}{\rho_{\rm sup}} \right) - \frac{4 \pi}{\mpl^2} \rho \left( 1 - 2 \frac{\rho}{\rho_{\rm sup}} \right),
\end{align}
\end{subequations}
where, $\rho$ is the energy density of the background scalar field and $\rho_{\rm sup}$ is the maximum energy density that can be attained. Calculations of black hole entropy fix the value of $\rho_{\rm sup}\, =\, 0.41\mpl^4$ \cite{Meissner:2004ju}. As is evident, at $\rho\, =\, \rho_{\rm sup}$, the scale factor reaches a minimum, {\it i.e.}, the universe undergoes a bounce when energy density becomes maximum. The evolution of the scalar field is governed by 
\begin{equation}\label{eqn:scalarfield}
    \ddot\phi\, +\, 3\,H\,\dot\phi\, +\, \frac{{\rm d}\,V(\phi)}{{\rm d}\phi}\,=\,0.
\end{equation}
Subsequent evolution after the bounce quickly decreases the energy density and makes the terms in Eqns. (\ref{eqn:lqc}) which are nonlinear in energy density, irrelevant, and these equations will reduce to the familiar Friedmann-Raychaudhuri equations in general relativity. 
Hence, if the scalar field is governed by a suitable potential, then slow roll inflation can set in some time after the bounce \cite{Ashtekar:2009mm, Ashtekar:2011rm, Bolliet:2017czc, Bhardwaj:2018omt}. 
\par 
In this work, we will consider two potentials for the scalar field, namely, the quadratic potential \cite{Linde:1983gd} and the Starobinsky potential \cite{Starobinsky:1980te}. The former is chosen because of its simplicity, and the latter for being one of the best-known models of slow-roll inflation \cite{Planck:2018jri}. The quadratic potential is given by 
\begin{equation}
V(\phi)\,=\,\frac{1}{2}m^2\phi^2,
\end{equation}
where we work with mass $m\,=\,1.353\times 10^{-6}\, \mpl$ that fits the observations from Planck \cite{Planck:2015fie}. 
The Starobinsky potential is given by 
\begin{equation}
V(\phi)\,=\, \frac{3\,m^2}{32\,\pi}\,\mpl^2\,\left( 1\,-\,e^{-\sqrt{\frac{16\,\pi}{3}}\frac{\phi}{\,\mpl}} \right)^2.
\end{equation}
We shall work with mass, $m\,=\, 2.676\times 10^{-6}\, \mpl$. 
\begin{figure}
    \centering
    \includegraphics[scale=0.38]{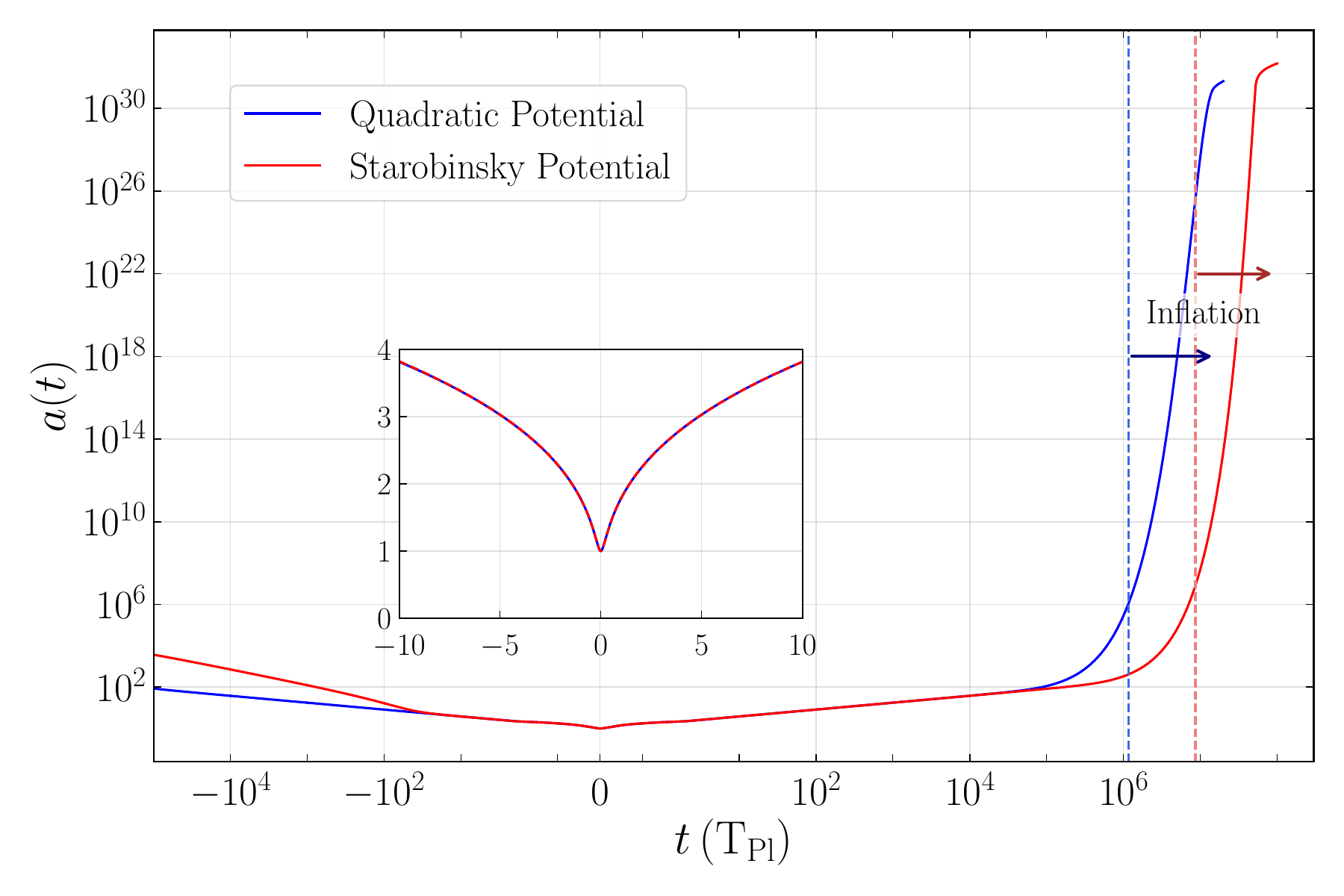}\\
    \begin{tabular}{cc}
         \includegraphics[scale=0.253]{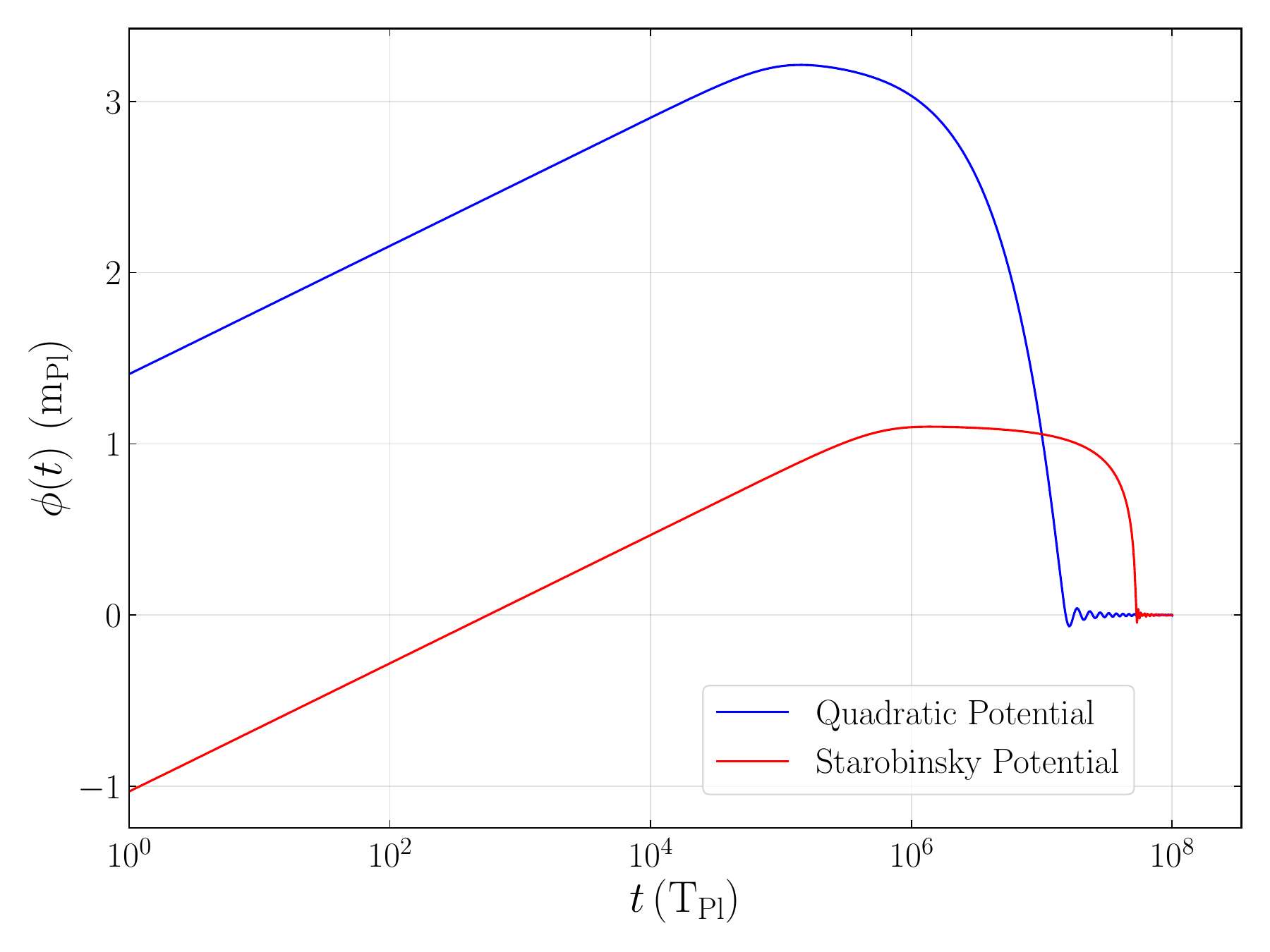}&  
         \includegraphics[scale=0.287]{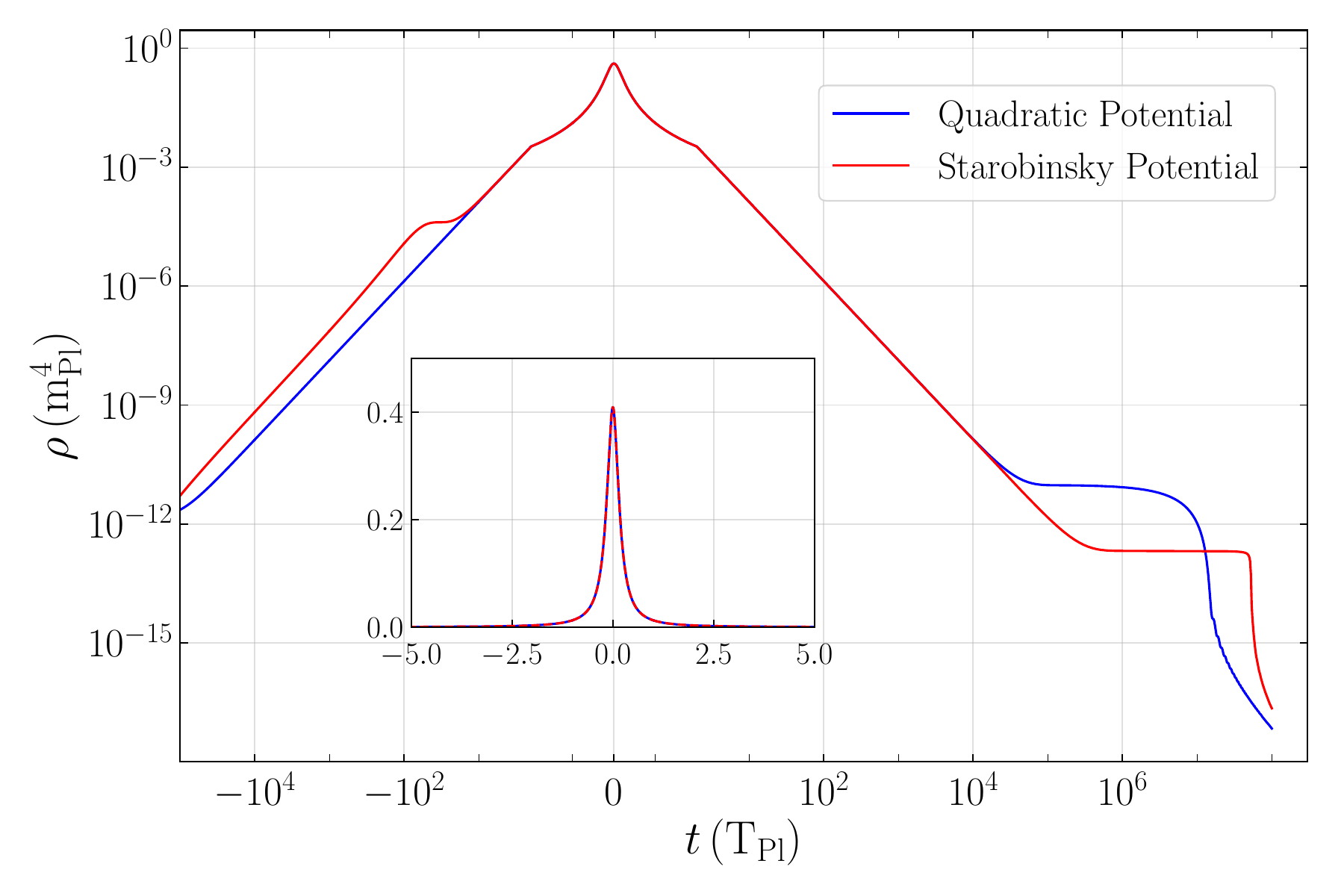}
    \end{tabular}
    \caption{\label{fig:background} Behaviour of scale factor (top), scalar field (bottom left), and energy density of scalar field (bottom right) in LQC for the initial conditions described in the text for the two potentials. From the plot of $a$, note that the epoch of inflation is preceded by a bounce, where an expanding phase transitions to a contracting phase. From the plot of $\phi$ we see that the scalar field reaches a maximum value sometime after the bounce and then decreases with time. For the scenario considered here, inflation is achieved during this epoch. From the plot of $\rho$, we see that energy density is maximum at the bounce. Further, we see from the inset that the behaviour of the scale factor near the bounce is independent of the form of the potential. }
    
\end{figure}
\begin{figure}
\end{figure}    
\par 
Loop quantum cosmology has another free parameter, namely the value of the scalar field at the bounce, denoted as $\phi_b$. For a given potential, the value of the scalar field at bounce determines the amount of expansion after the bounce till the time at which the pivot scale $k_\star\,=\, 0.002 {\rm Mpc}^{-1}$ leaves the horizon \cite{Agullo:2015tca, PhysRevD.97.066021}. More expansion prior to the exit of the pivot scale implies that signals from the bounce will not be visible at $k_\star$. In other words, more preinflationary expansion redshifts any signatures of the bounce away from the observable wavelengths. Pivot scale corresponds to a multipole of $\ell \approx 30$ in the cosmic microwave background. In the CMB power spectrum, most of the departure from near scale invariance occurs at multipoles of $\ell \lesssim 30$. We can put a lower bound on the value of $\phi_b$ by demanding that imprints of bounce are only visible at $k \lesssim k_\star$ ({\it i.e.}, at multipoles $\ell \lesssim 30$). This corresponds, roughly, to setting $k_\star\,\approx 3\,k_{\rm LQC}$, where $k_{\rm LQC}\,=\, a(t_B)\,\sqrt{8\,\pi\,\rho_B/\mpl^2}\,\approx\, 3.21 \mpl$. Following this approach, we have set $\phi_b\,= 1.014 \mpl$ in the case of quadratic potential and $\phi_b\,= -1.4243\, \mpl$ for Starobinsky potential. The velocity of the scalar field has been set to $\dot \phi_b\, =\, \sqrt{2\left(\rho_{\rm sup}\,-\,V(\phi_b)\right)}$. 
Such an initial condition will lead to about $14.00$ e-folds of expansion from bounce till exit of pivot scale and $56.88$ e-folds of expansion during slow roll for the case of quadratic potential. 
For the Starobinsky potential, this corresponds to a preinflationary expansion of about $15.86$ e-folds and $55.74$ e-folds since the exit of the pivot scale. 
\par 
The dynamics of background quantities $a$, $\phi$, and energy density $\rho$ in LQC for the two potentials with the initial conditions mentioned above are given in figure \ref{fig:background}. The figure illustrates that in LQC, the scale factor decreases initially, reaches a minimum value, undergoes a bounce, and then experiences inflationary expansion. From the plot of $\phi$, we see that, for the initial conditions we have considered, after the bounce, the kinetic energy of the scalar field is spent in climbing up the potential. After a while field reaches a maximum and starts rolling down. Slow-roll inflation sets in as the field begins to roll down. Inflation ends as the inflaton approaches the bottom of its potential. The scalar field then undergoes damped oscillations about that minimum. The figure also illustrates the behaviour of the energy density of the scalar field as a function of time. From the plots of $a$ and $\rho$, it is clear that energy density is maximum and equal to $\rho_{\rm sup}$ at the bounce. The energy density then quickly decays, and the quantum effects become negligible. The nearly flat behavior of the energy density indicates the epoch during which inflation occurs. Insets of scale factor and the energy density of the scalar field show that their behaviour is independent of the potential near the bounce.

\section{Numerical evolution of magnetic fields in LQC\label{sec:4}}
\begin{figure}
    \begin{tabular}{cc}
         \includegraphics[width=0.47\textwidth]{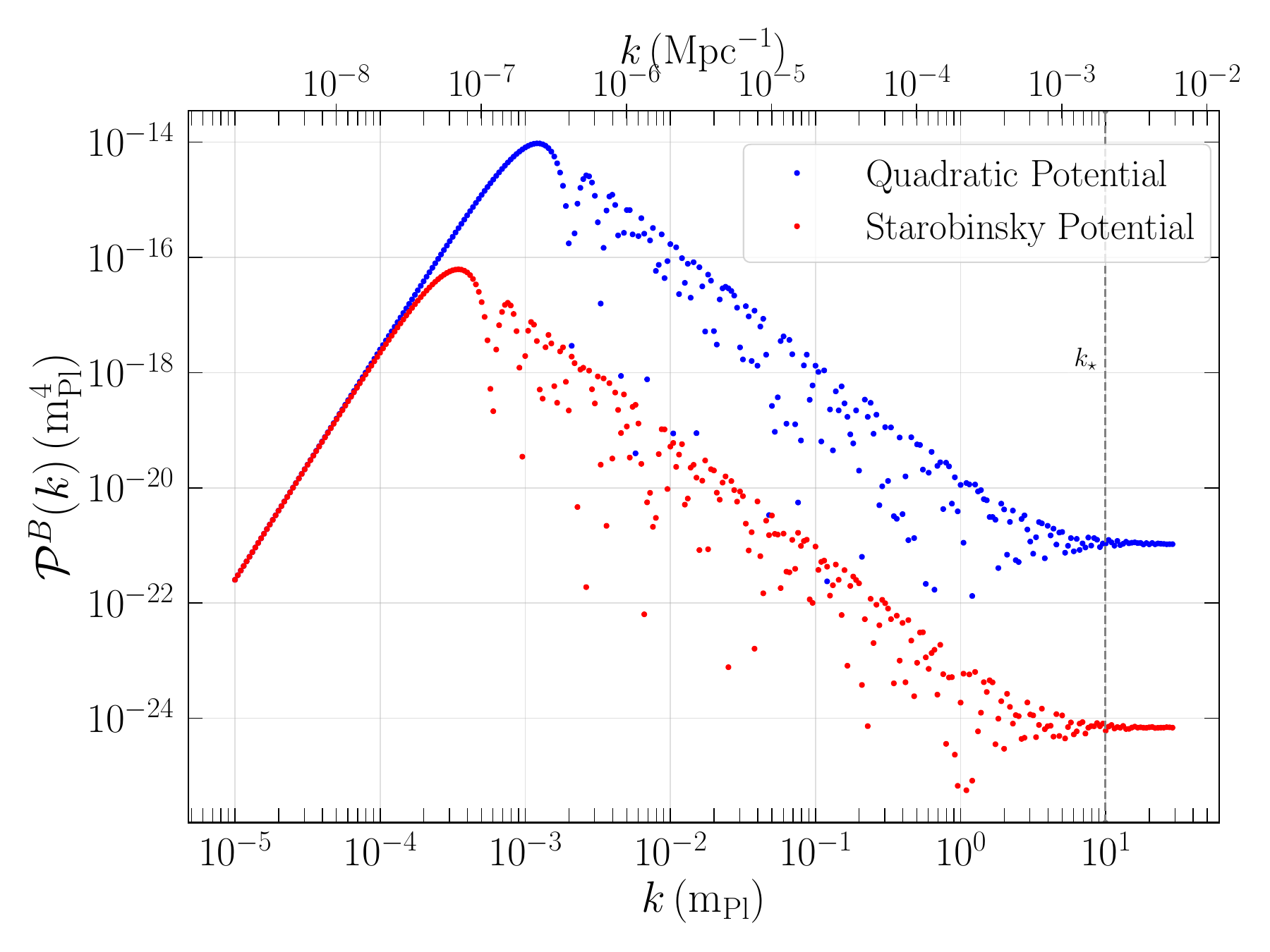}&  
         \includegraphics[width=0.47\textwidth]{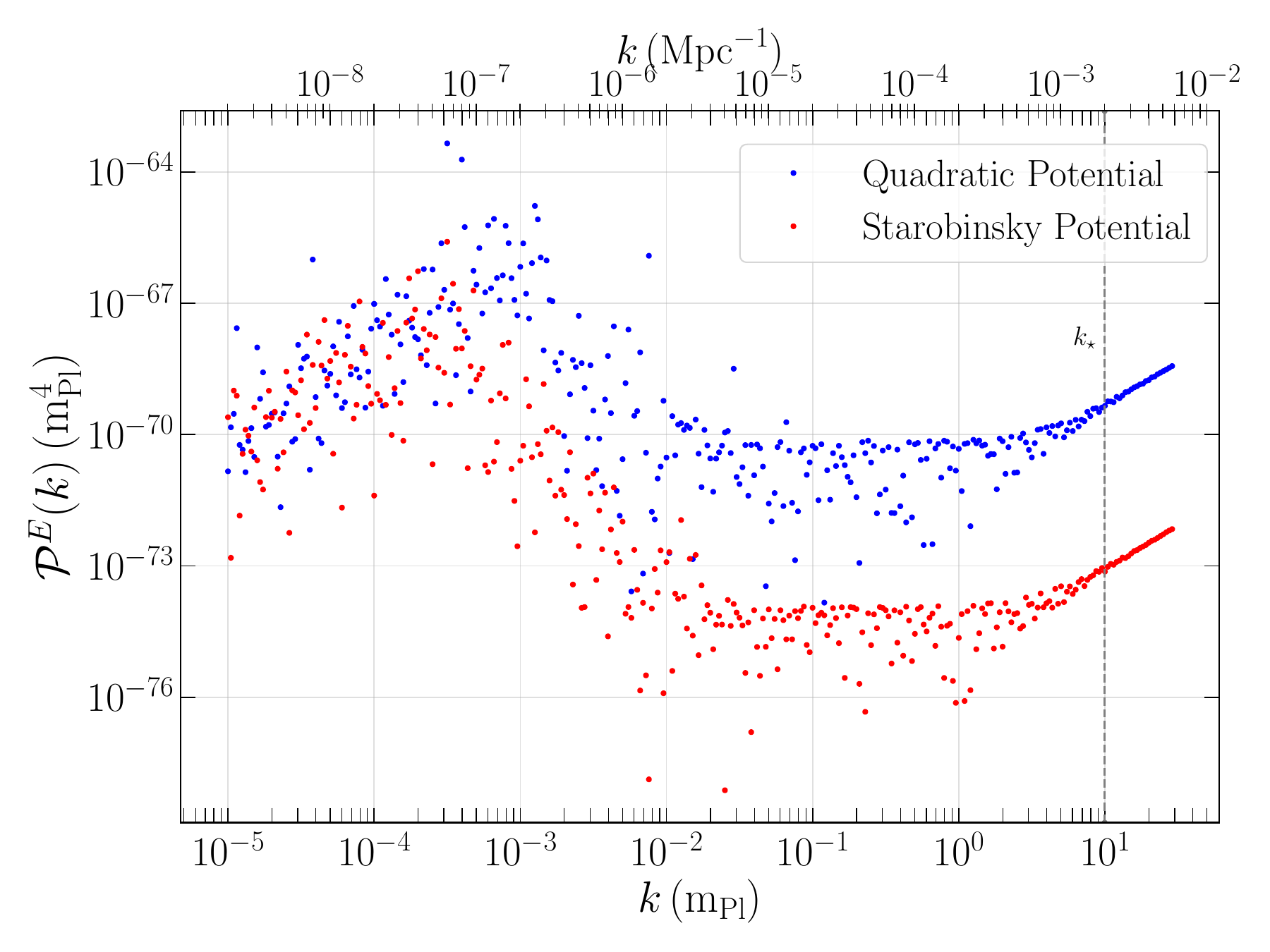}
    \end{tabular}
    \caption{\label{fig:PBPE} Plot of magnetic (left) and electric (right) power spectra generated in LQC. We have imposed the Minkowski initial condition at the bounce. We see that the power spectra of magnetic and electric fields generated in LQC are scale-dependent. Though magnetic power spectra at large wave numbers are scale invariant, intermediate wave numbers behave as $k^{-2}$ and infrared modes as $k^4$. 
    The magnetic spectrum scales as $1/k$ while transitioning between intermediate and large $k$.
    Electric power spectra have a $k^2$ behaviour at large wave numbers. For smaller wave numbers, the average spectra has a $k$ dependence, for even smaller wave numbers a scale-independent form, and for even smaller wave numbers the spectra has a $k^{-4}$ dependence. The electric spectra behave as $k^3$ at infrared scales.}
\end{figure}

We consider the vector potential as a quantum test field evolving on the quantum background of LQC. Numerically, the evolution of the vector potential in LQC is implemented in two stages. Firstly, we solve differential equations (\ref{eqn:lqc}, \ref{eqn:scalarfield}) and obtain the time evolution of background quantities.  Secondly, we evolve the vector potential on top of this background. The evolution of the magnetic field is governed by Eqn. (\ref{eqn:Ai}). For computing the power spectrum of magnetic fields, one needs to evolve the Fourier mode $A(t,k)$. The physics behind the evolution of the Fourier mode is more transparent if we express the equation of motion in terms of the variable ${\cal A}_k(\eta)\,=\,a(\eta)\,f(\eta)\,A(\eta,k)$ as 
\begin{equation}\label{eqn:Ak}
    {\cal A}_k''(\eta)\, +\, \left(\,k^2\,-\,\frac{f''(\eta)}{f(\eta)}\right){\cal A}_k(\eta)\,=\,0.
\end{equation}
Note that this is an equation of motion of a time-dependent harmonic oscillator. The evolution of the mode, hence, depends on the form of the coupling function. If the coupling function $f\,=\,1$, then the mode just oscillates and does not grow. Inspired from previous literature \cite{Martin_2008}, we shall consider the following form of coupling function,
\begin{equation}\label{eqn:fa}
    f(\eta)\, =\, \left(\frac{a(\eta)}{a(\eta_e)}\right)^n,
\end{equation}
with $n$ set to two. The value of $n$ has been chosen to be two, as it leads to a nearly scale-invariant spectrum with no backreaction in the case of slow-roll inflation. The evolution of the vector potential will also depend on the initial conditions of the vector potential. We shall impose Minkowski initial conditions, {\it viz.}, ${\cal A}_k\,=\, {\rm e}^{-i\,k\,\eta}/\sqrt{2\,k}$, at the bounce. We then numerically solve the equation of motion of the vector potential Eqn. (\ref{eqn:Ak}) from the bounce till the end of inflation. The power spectrum of the magnetic and electric field can then be computed using Eqns. (\ref{eqn:PB},  \ref{eqn:PE}) at the end of inflation. We have implemented this numerical procedure using \textit{Mathematica} \cite{Mathematica}. The plots of power spectra of magnetic and electric fields generated in LQC are given in figure \ref{fig:PBPE}. 

\section{Analysis of magnetogenesis in LQC \label{sec:5}}
We shall now try to understand various aspects that determine the scale dependence of the magnetic power spectra and their implications.
\begin{figure}
\includegraphics[width=0.7\textwidth]{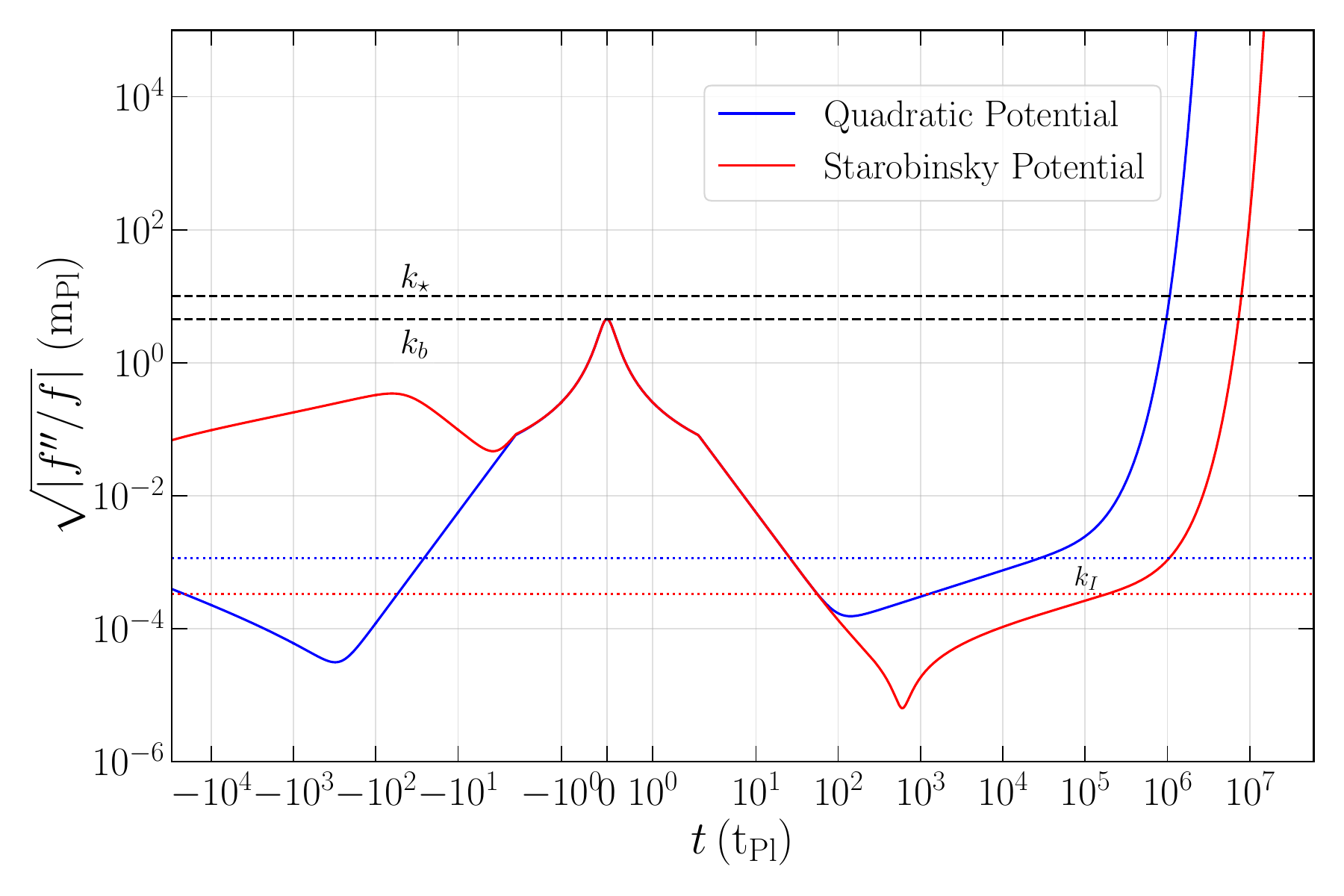}
\caption{\label{fig:fppbyf} Plot of comoving wave numbers (horizontal dashed and dotted lines) and $\sqrt{|f''/f|}$, for both quadratic potential (blue) and Starobinsky potential (red), as a function of time. The peak of $\sqrt{|f''/f|}$ at the bounce sets a scale $k_b\,=\, \left(\sqrt{|f''/f|}\right)\biggl\vert_{t\,=\,0}$. 
Another relevant scale in the problem is $k_I$, which refers to the largest wave number that becomes smaller than $\sqrt{|f''/f|}$ just before the onset of inflation. Modes with $k>>k_b$ will only be excited when they cross $\sqrt{|f''/f|}$ during inflation. Hence, their spectra will be similar to that generated in a slow-roll inflationary scenario. 
Modes with $k_I \lesssim\,k\, \lesssim k_b$ cross $\sqrt{|f''/f|}$ during the bounce and hence are in an excited state at the onset of inflation. The spectra of these modes will be scale-dependent. Modes with $k<<k_I$ are largely not excited throughout the evolution. } 
\end{figure}
\subsection{Origin of scale dependence}\label{sec:5A}
The origin of the scale dependence of the vector potential can be understood by studying Eqn. (\ref{eqn:Ak}). From this equation, we can see that, when $k^2 >> f''/f$ the modes are simply oscillatory and when $k^2 << f''/f$, the modes evolve as ${\cal A}_k\, \propto\, f$. Thus, the evolution of the Fourier mode of the vector potential can be understood by comparing the term $\sqrt{|f''/f|}$ with the wave number of perturbations. We plot $\sqrt{|f''/f|}$ as a function of time, for both potentials of scalar fields in figure \ref{fig:fppbyf}. We have also plotted comoving wave numbers, as horizontal dashed and dotted lines, for comparison. Here, $k_b$ is a scale set by the peak of the function $\sqrt{|f''/f|}$ at the bounce. Note that $k_b \approx k_{LQC}$ for the coupling function we have chosen. If we impose Minkowski initial conditions at the bounce,  we can see from the plot that modes $k \lesssim k_b$ are excited during the bounce and hence are not in the Minkowski vacuum at the onset of inflation. Whereas, modes $k >> k_b$ are not excited during the bounce and hence remain in the Minkowski vacuum at the onset of inflation. Thus, the behavior of modes with $k >> k_b$ will be similar to the behavior of these modes in slow-roll inflation. For the coupling function Eqn. (\ref{eqn:fa}), with $n=2$, we expect nearly scale-invariant magnetic power spectra for modes $k>>k_b$ and scale-dependent behaviour for smaller wave numbers. 
From figure \ref{fig:PBPE} we see that the magnetic spectrum of modes $k>>k_b\simeq k_\star/3$ is nearly scale-invariant, whereas intermediate modes with $k_I\lesssim k < k_b$ behave on average as $k^{-2}$.  Here, $k_I$ is the wave number which becomes smaller than $\sqrt{|f''/f|}$, {\it i.e.}, becomes superhorizon, just before the onset of inflation. Power spectra briefly behave as $1/k$ for wave numbers around $k_b$, {\it i.e.}, as the behaviour transitions from nearly scale invariant to a $k^{-2}$ behaviour.
wave numbers with $k<<k_I$ satisfy the condition $k^2<<f''/f$ throughout its evolution and hence for these modes, ${\cal A}_k\, \propto\, f$. Thus, for $k<<k_I$, we see from Eqn. (\ref{eqn:PB}) that, for $n\,=\,2$, the magnetic spectra behaves as $k^4$. This behaviour of infrared modes is also evident from the figure \ref{fig:PBPE}. 
\par 
Similarly, our simulations also show that the electric power spectra are proportional to $k^2$ as in the case of slow-roll inflation for large wave numbers. Average behaviour of electric power spectrum transitions to a $k$ dependence, then to a scale-invariant behaviour, and to $k^{-4}$ behaviour as we move to smaller and smaller scales. At far infrared scales,  $k<<k_I$, the electric spectra behave as $k^3$. The behaviour of far infrared modes can be understood as follows. These modes always satisfy $k^2 << f''/f$. Hence, these modes behave as ${\cal A}_k\,\propto C_1\,f\, + C_2/(a\,f)$, where $C_1$ and $C_2$ are constants. From Eqn. (\ref{eqn:PE}), we see that the contribution to the electric power spectrum arises from the decaying part of the mode. Upon substituting the decaying modes in Eqn. (\ref{eqn:PE}), we obtain electric power spectrum for these modes to be proportional to $k^3$.
\par 
\begin{figure}
    \begin{tabular}{cc}
         \includegraphics[width=0.47\textwidth]{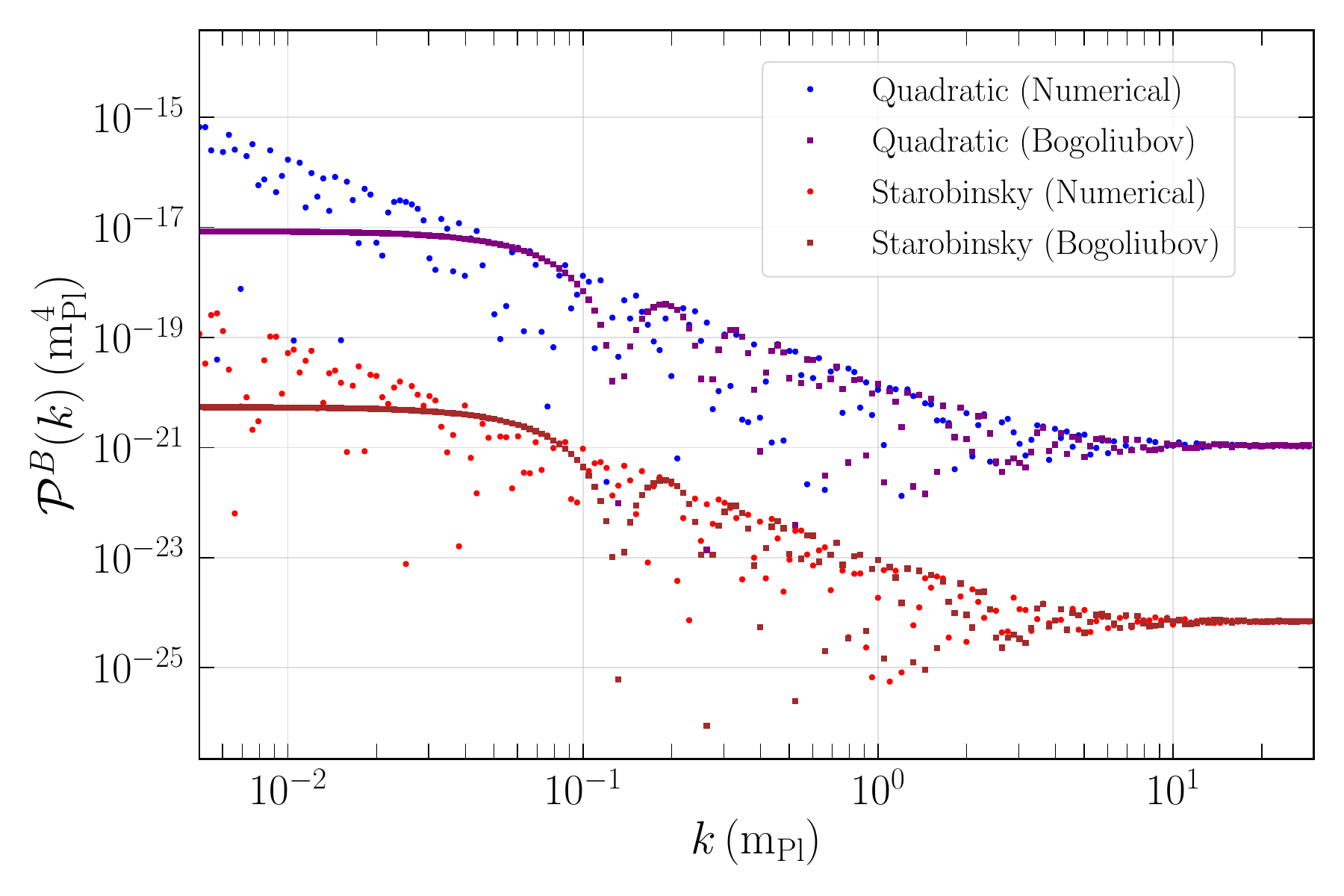}&  
         \includegraphics[width=0.47\textwidth]{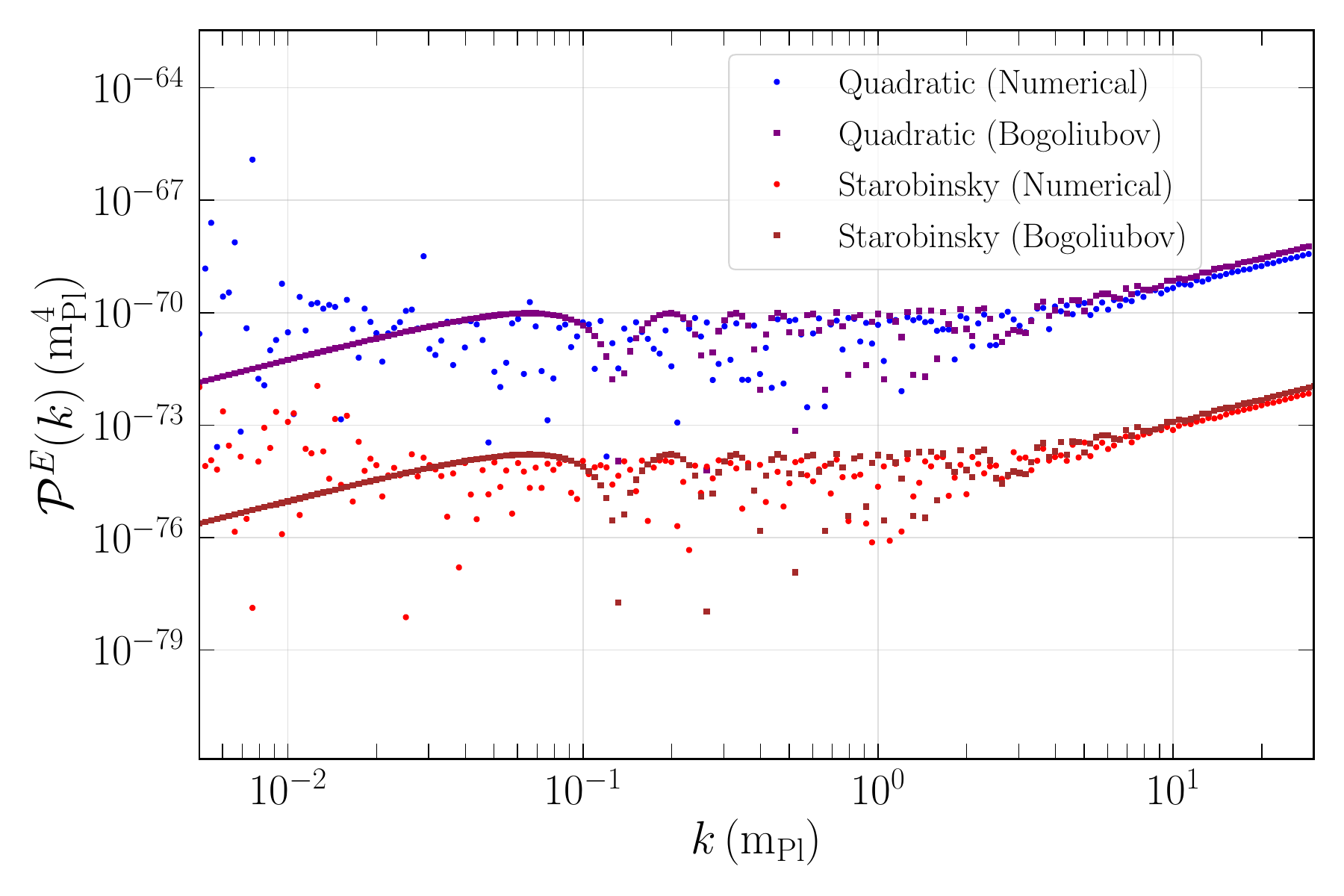}
    \end{tabular}
\caption{\label{fig:ps-bogo} Comparison of magnetic and electric power spectra obtained using Eqn. (\ref{eqn:ps-bogo}) with exact numerical results. The two calculations agree, indicating that bounce excites the state of perturbations from their initial Minkowski vacuum. The two results are not expected to agree at infrared wave numbers as those modes are not adiabatic at $t_{\rm mink}$ and hence we cannot approximate their behaviour by using Eqn. (\ref{eqn:ps-bogo}).}
\end{figure}
We shall now try to understand, in more detail, the effect of bounce on the evolution of modes. For this, let us consider a time $t_{\rm mink}$ after the bounce, when most of the observable modes are adiabatic, {\it i.e.}, satisfy the condition $k^2\, >>\, f''/f$. Suppose ${\cal A}_k(t)$ are solutions to Eqn. (\ref{eqn:Ak}), if Minkowski initial conditions are imposed at the bounce and ${\cal A}^{\rm mink}_k(t)$ are solutions obtained upon imposing initial conditions at $t_{\rm mink}$, then we can express ${\cal A}_k(t)$ in terms of ${\cal A}^{\rm mink}_k(t)$ and its complex conjugate as 
\begin{equation}
    {\cal A}_k(t)\, =\, \alpha_k\,{\cal A}^{\rm mink}_k(t)\, +\, \beta_k\, {\cal A}^{\rm mink^*}_k(t),
\end{equation}
where $\alpha_k$ and $\beta_k$ are Bogoliubov's coefficients given by 
\begin{eqnarray}
\alpha_k\,=\, i\,a(t)\,\left( {\cal A}_k^{\rm mink^*}\, \dot{\cal A}_k\, -\, \dot{\cal A}_k^{\rm mink^*}\, {\cal A}_k\right),\nonumber\\
\beta_k\,=\, -i\,a(t)\,\left( {\cal A}_k^{\rm mink}\, \dot{\cal A}_k\, -\, \dot{\cal A}_k^{\rm mink}\, {\cal A}_k\right).\nonumber
\end{eqnarray}
They satisfy $|\alpha_k|^2\,-\,|\beta_k|^2\,=\,1$. We can use the above expressions to express the magnetic and electric power spectra of ${\cal A}_k$ in terms of ${\cal A}_k^{\rm mink}$. Since the spectra for ${\cal A}_k^{\rm mink}$ are equal to that obtained in slow-roll for large enough wave numbers, we can approximate, up to a phase, magnetic and electric spectra obtained in LQC in terms of Bogoliubov coefficients as 
\begin{eqnarray}\label{eqn:ps-bogo}
    {\cal P}^{B,\,E}(k)\, =\, \left|\alpha_k\, +\, \beta_k\right|^2\, {\cal P}^{B,\,E}_{SR}(k).
\end{eqnarray}
The spectra obtained in slow-roll inflation (see, for instance, \cite{Sagarika_2022}) are,
\begin{eqnarray}
\frac{{\cal P}^{B}_{SR}(k)}{m_{Pl}^4}\,\simeq\, \frac{9\,\pi^2}{16}\, (A_s\,r)^2,\\
\frac{{\cal P}^{E}_{SR}(k)}{m_{Pl}^4}\,\simeq\, \frac{\pi^2}{16}\, (A_s\,r)^2\,(k/k_e)^2,
\end{eqnarray}
where $A_s$ refers to the amplitude of primordial scalar perturbations, $r$ is the tensor-to-scalar ratio, and $k_e$ is the mode that exits the horizon at the end of inflation. A comparison of power spectra obtained from Eqn. (\ref{eqn:ps-bogo}) with exact numerical computation is given in figure \ref{fig:ps-bogo}.
The good agreement between the two curves shows that bounce preceding inflation modifies the initial state of modes with smaller wave numbers from the Minkowski initial states. Further, in agreement with numerical calculations, the above expressions also explain that the Starobinsky model will lead to a lower power of the magnetic field due to its lower tensor-to-scalar ratio. 

\subsection{Effect of imposing initial conditions at different time}\label{sec:5B}
\begin{figure}
    \centering
    \includegraphics[scale=0.4]{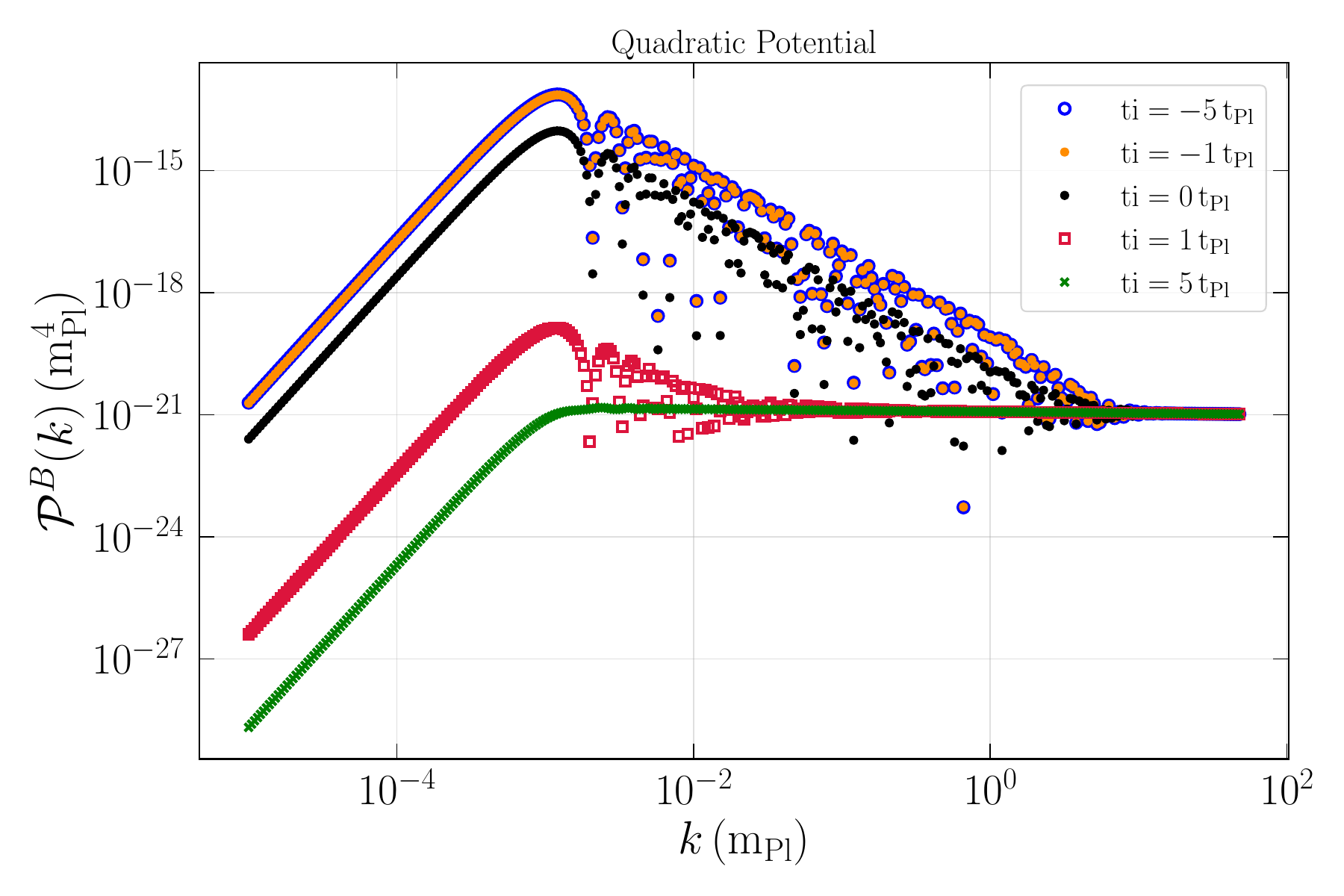}
    \caption{\label{fig:PBITQP} Magnetic power spectrum generated in LQC with the quadratic potential when Minkowski initial conditions are imposed at different times.}
\end{figure}

\begin{figure}
    \centering
    \includegraphics[scale=0.4]{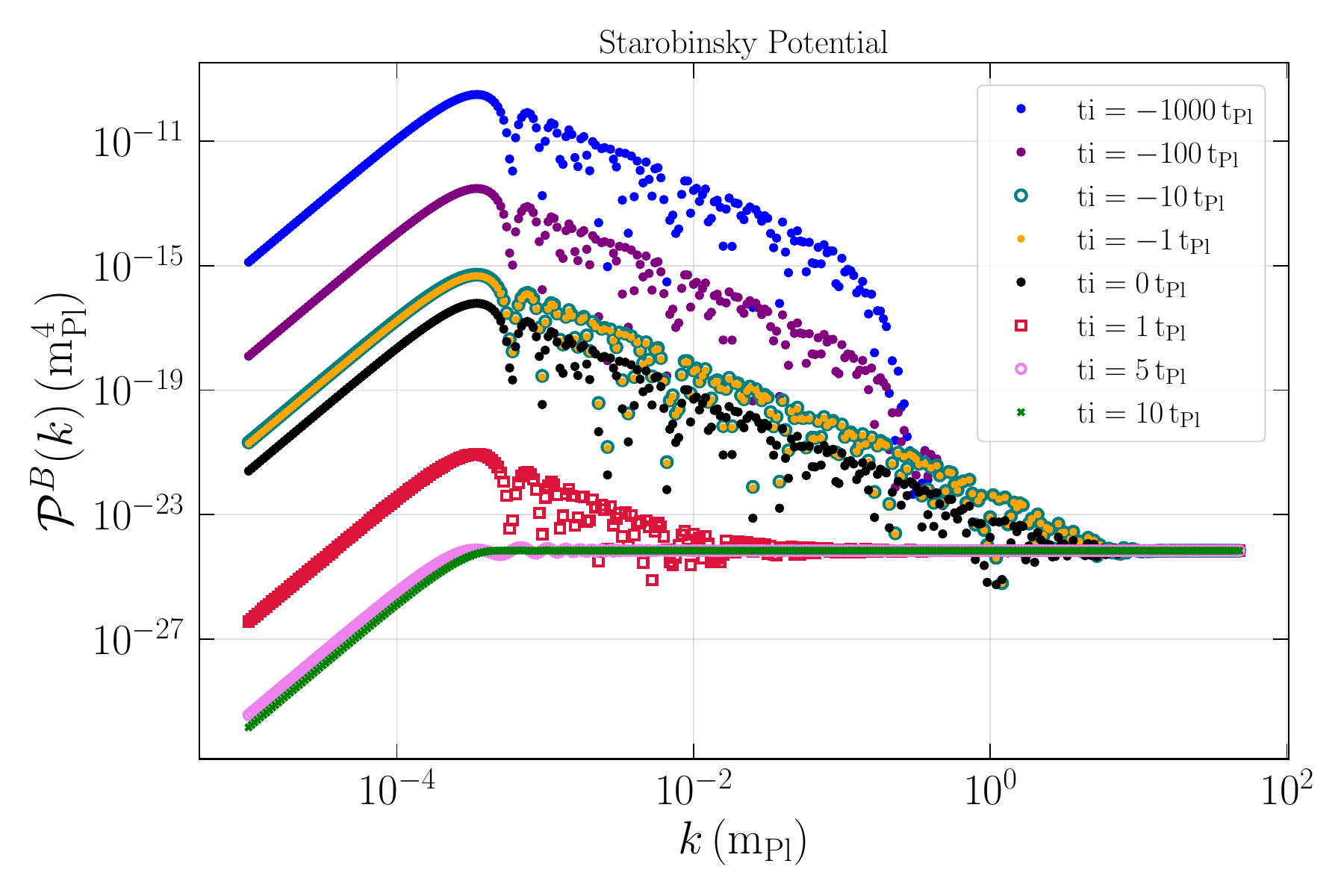}
    \caption{\label{fig:PBITSM}Magnetic power spectrum generated in LQC with the Starobinsky potential when Minkowski initial conditions are imposed at different times.}
\end{figure}

In obtaining figures \ref{fig:PBPE} and \ref{fig:ps-bogo} we imposed Minkowski initial condition  at the bounce, {\it i.e.}, at $t\,=\,0\, \tpl$. In this section, we study the dependence of magnetic power spectra on the time at which Minkowski initial conditions are imposed. Figures \ref{fig:PBITQP} and \ref{fig:PBITSM} contain graphs of magnetic power spectra generated in LQC with quadratic and Starobinsky models, respectively, upon imposing Minkowski initial conditions at different times. The spectra in each case have been computed after modes become $k<< \sqrt{|f''/f|}$, {\it i.e.}, superhorizon, during inflation. Note that this is equivalent to computing the magnetic spectrum at the end of inflation. This is because the magnetic power spectrum remains constant after the mode becomes superhorizon for the case of the coupling function with $n=2$.
\par 
Figure \ref{fig:PBITQP} corresponds to the magnetic spectra generated in the case of the quadratic potential. We find that when initial conditions are imposed before the bounce, the obtained spectra are slightly higher than the one obtained upon imposing initial conditions at the bounce. However, we find that the spectra obtained with initial conditions imposed at any time prior to the bounce are largely similar. This may be because of the sharp drop in the amplitude of $\sqrt{|f''/f|}$ before the bounce (see figure \ref{fig:fppbyf}). When initial conditions are imposed after the bounce, we see that the scale dependence decreases a lot and the spectra become similar to that obtained during inflation. This can again be attributed to the sharp drop in the amplitude of $\sqrt{|f''/f|}$ after the bounce. The scale dependence of infrared modes, {\it i.e.}, $k\,<\,k_I$, remains the same regardless of the time at which the initial condition is imposed. As expected, they display a $k^4$ dependence since, as explained in the previous section, these modes are at most of the time, superhorizon.
\par
Figure \ref{fig:PBITSM} plots magnetic power spectra generated in the Starobinsky model when the initial conditions are imposed at different times. As in the case of the quadratic potential, we impose the Minkowski initial condition, and the spectra are computed when modes are superhorizon. 
The dependence of power spectra generated in the case of the Starobinsky model is more involved. This is expected and can be understood from the behaviour of $\sqrt{|f''/f|}$ given in figure \ref{fig:fppbyf}. As in the case of quadratic potential, the spectra generated when initial conditions are imposed before the bounce are larger than that obtained when initial condition is imposed at the bounce. However, unlike in the case of quadratic potential, the amplitude of the spectrum for modes $k_I<k<<k_b$ depends strongly on the time at which initial conditions are imposed. We find that the earlier the initial time, the larger the amplitude of the spectrum for these scales. This behaviour may be attributed to the presence of an additional bump in $\sqrt{|f''/f|}$, see figure \ref{fig:fppbyf}, for the Starobinsky model. The extra bump excites the modes even before they are excited during the bounce. The behaviour of $\sqrt{|f''/f|}$ after the bounce is similar in both potentials. Hence, the dependence of the power spectrum on the initial time, if the initial conditions are imposed after the bounce, is similar to that obtained in the case of the quadratic potential. The scale dependence of the power spectrum decreases a lot and becomes largely similar to that obtained in the context of slow-roll inflation if the initial time is farther from the bounce. As in the case of the quadratic potential, the spectra of infrared modes behave as $k^4$ irrespective of the time at which the initial condition is imposed.

\par 
It is interesting to note that imposing Minkowski initial conditions at different times is equivalent to imposing excited initial conditions at the bounce. This is because the Minkowski initial condition imposed at a time $t$ can be evolved to the time of the bounce to get an excited state. The magnetic spectra obtained upon imposing this excited state as an initial condition at the bounce will be the same as the one obtained upon imposing the Minkowski initial condition at time $t$. 
\subsection{Some remarks on equivalent forms of coupling function}\label{sec:5C}
Till now, we considered a coupling function given in Eqn. (\ref{eqn:fa}). It would be interesting to investigate whether we can arrive at coupling functions in terms of scalar field or other background quantities, such as curvature. In this subsection, we will investigate the viability of some such proposals in the context of LQC.
\begin{figure}
\includegraphics[width=0.7\textwidth]{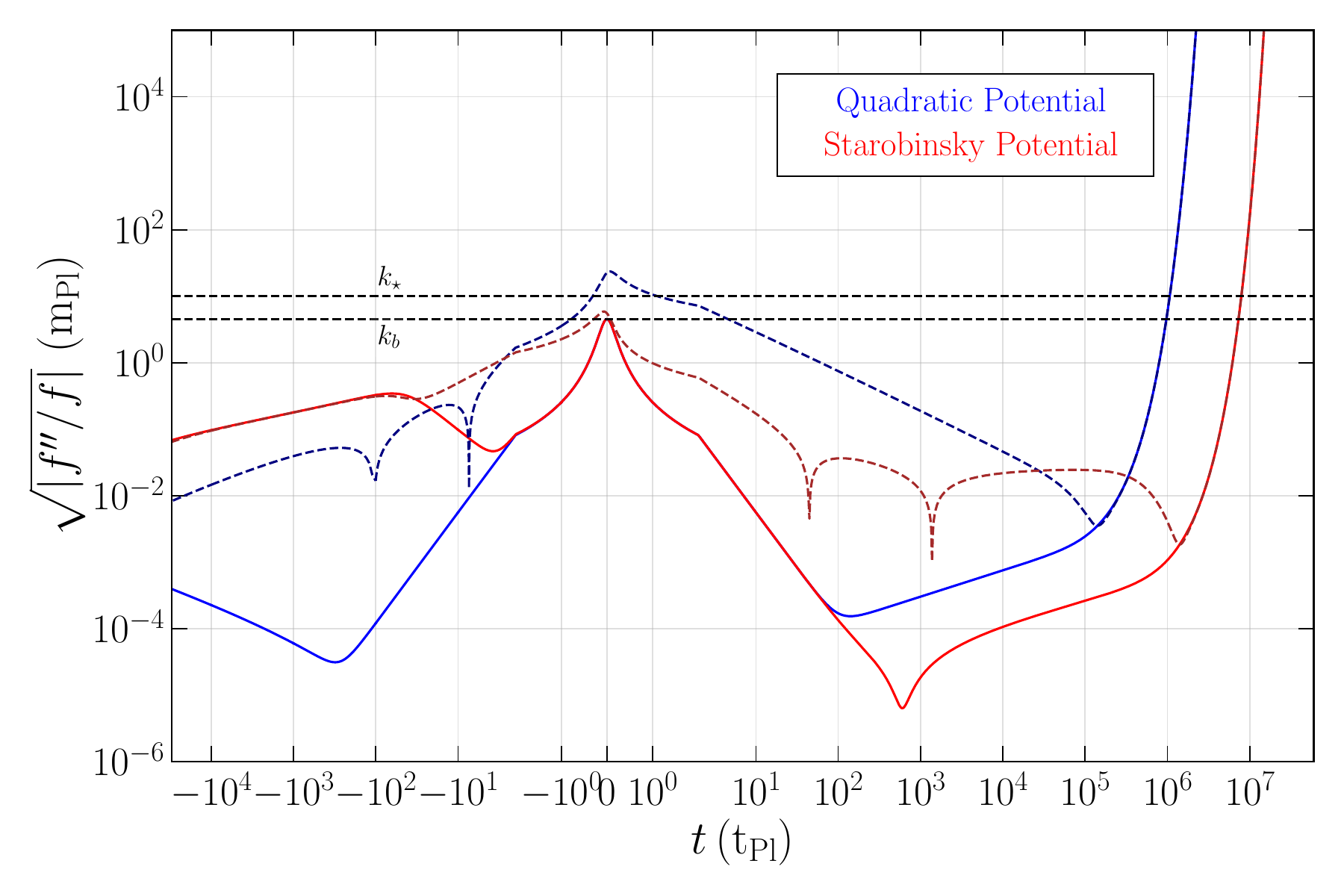}
\caption{\label{fig:fppbyfphi} Plot of comoving wave numbers (black dashed lines) and $\sqrt{|f''/f|}$, for both quadratic potential (blue) and Starobinsky potential (red), as a function of time. Solid curves represent $\sqrt{|f''/f|}$ corresponding to Eqn. (\ref{eqn:fa}) and 
dashed curves corresponding to Eqns. (\ref{eqn:fphiqp}) and (\ref{eqn:fphisp}). A larger value of $\sqrt{|f''/f|}$ for the dashed curves near the bounce leads to a larger power spectrum. } 
\end{figure}
\par 
First, we shall consider a coupling function in terms of the scalar field, which is equivalent to Eqn. (\ref{eqn:fa}) during slow-roll (see, for instance, \cite{Sagarika_2022}). Such a function can be set up as follows. In the case of the quadratic potential, 
\begin{equation}
    \phi^2(N)\, \simeq\, \phi_e^2\,+\,\frac{(N_e\,-\,N)}{2\,\pi}\,\mpl^2,
\end{equation}
with $\phi_e\,\approx\,1/\sqrt{4\, \pi}\,\mpl$. 
Using the above expression, we can construct the coupling function similar to Eqn. \ref{eqn:fa} as 
\begin{equation}\label{eqn:fphiqp}
    f(\eta)\,=\, \exp\left[ -\frac{2\,\pi\, n}{\mpl^2}\,( \phi^2\,-\,\phi_e^2) \right].
\end{equation}
Similarly, in the case of the Starobinsky model, the field value is related to e-folds as
\begin{eqnarray}
    N\,-\,N_e\, \simeq\, -\frac{3}{4}\, \biggl[\,\exp\left(\sqrt{\frac{16\,\pi}{3}}\frac{\phi}{\mpl} \right)\,-\, \exp\left(\sqrt{\frac{16\,\pi}{3}}\frac{\phi_e}{\mpl} \right)\,-\,\sqrt{\frac{16\,\pi}{3}}\,\frac{(\phi\,-\,\phi_e)}{\mpl} \,\biggr],
\end{eqnarray}
where $\phi_e\, =\, \sqrt{\frac{3}{16\,\pi}}\,\ln \left(\,1\,+\,2/\sqrt{3} \right)\, \mpl$. This expression can be used to construct a coupling function equivalent to Eqn. (\ref{eqn:fa}), 
\begin{eqnarray}\label{eqn:fphisp}
    f(\eta)\, &=&\, \exp\biggl\{\,-\frac{3\,n}{4}\, \biggl[\,\exp\left(\sqrt{\frac{16\,\pi}{3}}\frac{\phi}{\mpl} \right)\,-\, \exp\left(\sqrt{\frac{16\,\pi}{3}}\frac{\phi_e}{\mpl} \right)\,\nonumber\\
    &&-\,\sqrt{\frac{16\,\pi}{3}}\,\frac{(\phi\,-\,\phi_e)}{\mpl} \,\biggr] \biggr\}.
\end{eqnarray}
Recall that we have set $n\,=\,2$.
\par
In figure \ref{fig:fppbyfphi}, we have compared $\sqrt{|f''/f|}$ corresponding to Eqn. (\ref{eqn:fa}) generated in the quadratic and the Starobinsky models, with that corresponding to Eqns. (\ref{eqn:fphiqp}) and (\ref{eqn:fphisp}). As expected, we see that though different $\sqrt{|f''/f|}$ matches in the slow-roll regime, they depart before the inflationary era. The amplitude of $\sqrt{|f''/f|}$ corresponding to Eqns. (\ref{eqn:fphiqp}) and (\ref{eqn:fphisp}) are very large closer to the bounce, and we find that such a large $\sqrt{|f''/f|}$ leads to a large magnetic spectrum. Such a large magnetic power spectrum will cause backreaction. This breaks our assumption that magnetic fields are test fields living on the FLRW background. Hence, coupling functions considered in Eqns. (\ref{eqn:fphiqp}) and (\ref{eqn:fphisp}) are not suitable for magnetogenesis in the context of LQC.
\par 
One may express the coupling function Eqn. (\ref{eqn:fa}) in terms of Ricci curvature as follows \cite{Turner_Widrow_1988, Bamba:2008ja}. In FLRW spacetime, Ricci scalar is  $R\,=\, 6\,\left(\, \ddot a/a\,+\,H^2\,\right)$. The equivalent coupling function can then be expressed as 
\begin{equation}\label{eqn:fRtrivial}
    f(R)\,=\, \biggl( \frac{R}{6\,H_e^2} \biggr)^{\alpha(N)},
\end{equation}
where $\alpha(N)\,=\,2(N\,-\,N_e)/\ln\,\left[ R/(6\,H_e^2)\right]$. One can easily verify that the above expression is the same as Eqn. (\ref{eqn:fa}). 
Similarly, we can also write the coupling function Eqn. (\ref{eqn:fa}) in terms of the scalar field as 
\begin{equation}\label{eqn:fphitrivial}
    f(\phi)\,=\, \biggl( \frac{\phi(N)}{\mpl} \biggr)^{\alpha(N)},
\end{equation}
where $\alpha(N)\,=\,2(N\,-\,N_e)/\ln{\left(\phi(N)/\mpl\right)}$.  This too is equivalent to Eqn. (\ref{eqn:fa}) in LQC. 
\subsection{Backreaction}\label{sec:5D}
One of the key assumptions that we have made is that the electromagnetic vector potential is a test field living on a background dominated by the scalar field. It is hence important to check whether this assumption holds true. In particular, we would like to understand whether the energy density of the electromagnetic field computed at the end of inflation is smaller than that of the scalar field, {\it i.e.}, 
\begin{equation}
    \rho^\phi\, >>\, \rho^B\,+\rho^E,
\end{equation}
where $\rho^\phi\,=\,\dot\phi^2/2\,+\,V(\phi)$. Note that $\rho^{\phi}$ is same as $\rho$ appearing in Eqns. (\ref{eqn:lqc}). The energy densities of magnetic and electric fields are given by Eqns. (\ref{eqn:rhoB}) and (\ref{eqn:rhoE}). They are related to the power spectra through Eqns. (\ref{eqn:PB}) and (\ref{eqn:PE}). Since the power spectrum of the electric field that is produced with the coupling function Eqn. (\ref{eqn:fa}) is much smaller than that of the magnetic field, see figure \ref{fig:PBPE}, it is sufficient to check whether $\rho^\phi\, >>\, \rho^B$. We numerically integrated over the magnetic power spectrum Eqn. (\ref{eqn:PB}) to obtain the corresponding energy density at the end of inflation. 
We find that, at the end of inflation, for the coupling function described by Eqn. (\ref{eqn:fa}), and when Minkowski initial conditions are imposed at the bounce
\begin{eqnarray}
    \frac{\rho^B (t_e)}{\rho^\phi  (t_e)}\,&=&\, 
    \begin{cases}
   \;\; 0.16\;\;\; ({\rm Quadratic \, potential})\\
    \;\;0.0011\;\;\; ({\rm Starobinsky \, potential}).
    \end{cases}                       
\end{eqnarray}
This calculation shows that, for the initial condition and coupling function considered here, the magnetic field can be considered as a test field for the case of the Starobinsky model. The energy density of the magnetic field, though smaller than that of the scalar field, may not be negligible for the case of a quadratic potential. 
We also computed the ratio of magnetic energy density to energy density of the scalar field for the excited states studied in section \ref{sec:5B}. We see that in the case of quadratic potential, the ratio of magnetic energy density to that of scalar field varies from $1.2\, <\, \rho^B/\rho^\phi\, <\, 2.4\times 10^{-7}$, where the higher ratio is obtained when Minkowski initial condition was imposed at $t\,=\,-5\,\tpl$ and lower ratio corresponds to an initial time of $t\,=\, 5\,\tpl$. For the Starobinsky potential, we find that we obtain significant backreaction when Minkowski initial conditions are imposed before the bounce. When Minkowski initial conditions are imposed after the bounce, say at $t\,=\,10\, \tpl$, $\rho^B/\rho^\phi\,=\, 1.7\,\times\,10^{-10}$. Since imposing Minkowski initial conditions at different times is equivalent to imposing excited (non-Minkowski) initial states at the bounce, our calculations show that the energy density of magnetic fields depends on the initial condition.  Thus, we see that for both models, there exist initial conditions for which backreaction is negligible. 
\subsection{Estimate of magnetic field present today}\label{sec:5E}
The electric field generated during inflation will be shorted out as soon as the Universe becomes conducting. Hence, there will not be any remnant large-scale electric field. However, there will be some amount of primordial magnetic field that will be observable today. We will now compute the value of the magnetic field generated in LQC that can be observed today. 
As in earlier sections, we shall assume a coupling function of the form Eqn. (\ref{eqn:fa}). After the end of inflation, the energy density of the magnetic field decays as the fourth power of the scale factor. Further, the magnetic field will be related to energy density as $\rho^B(t)\,=B^2(t)/2$. So primordial magnetic field existing today is related to the energy density of magnetic fields at the end of inflation as
\begin{equation} \label{eqn:B0}
B_0\,=\,\sqrt{2\,\rho^B(t_0)}\,=\,\sqrt{2\,\rho^B(t_e)}\,\left(\frac{a_e}{a_0}\right)^2\,.
\end{equation}
The amount of expansion between the end of inflation until today can be fixed by assuming that entropy given by $g_s\,T^3\,a^3$ is conserved during this period (see, for instance, \cite{Martin_2008}). Here, $g_s$ is the effective relativistic degrees of freedom that contribute to entropy, and $T$ is the temperature at an epoch with scale factor $a$. Then, the amount of expansion between the end of inflation till today is
\begin{equation}
    \frac{a_0}{a_e}\,=\,\biggl(\frac{g_{s,e}}{g_{s,0}}\biggr)^{1/3}\,\frac{T_e}{T_0}.
\end{equation}
We work with $g_{s,e}\,=\,106.75$, $g_{s,0}\,=\,3.36$ and $T_0\,=\,2.73\, K$. If we assume instantaneous reheating followed by a phase of radiation domination, the energy density at the end of inflation can be equated to the temperature $T_e$ as
\begin{equation}
\rho_I\,=\,\frac{3\,H_I^2}{8\pi}\,m_{\rm Pl}^2\,=\, g_{r,e}\,\frac{\pi^2}{30}\,T_e^4.
\end{equation}
If we assume that $g_{r,e}$, which is the effective number of relative degrees of freedom that contribute to the energy density of radiation, is the same as $g_{s,e}$, then we obtain 
\begin{equation}
T_e\,=\,\biggl(\frac{90\,H_I^2\,m_{\rm Pl}^2}{8\,\pi^3\,g_{s,e}}\biggr)^{1/4}.
\end{equation}
This in turn gives us
\begin{equation}
\frac{a_0}{a_e}\,=\,1.26\,\times\,10^{29}\,\biggl( \frac{H_I}{10^{-5}\,m_{\rm Pl}}\biggr)^{1/2}.
\end{equation}
We can substitute the above expression in \ref{eqn:B0} to obtain $B_0$. We find that for the quadratic potential, if Minkowski initial condition is imposed at the bounce, the strength of the primordial magnetic field existing today is $B_0\, =\, 0.94\, \mu{\rm G}$. For Starobinsky model, value of $B_0\, =\, 0.08\, \mu{\rm G}$. We also find that if Minkowski initial condition was imposed at $t\,=\,1\,\tpl$, $B_0\, = 3.7\,{\rm nG}$ and $B_0\, = \,0.3\,{\rm nG}$ in quadratic and Starobinsky models, respectively. 
\par
\begin{figure}
    \includegraphics[scale=0.4]{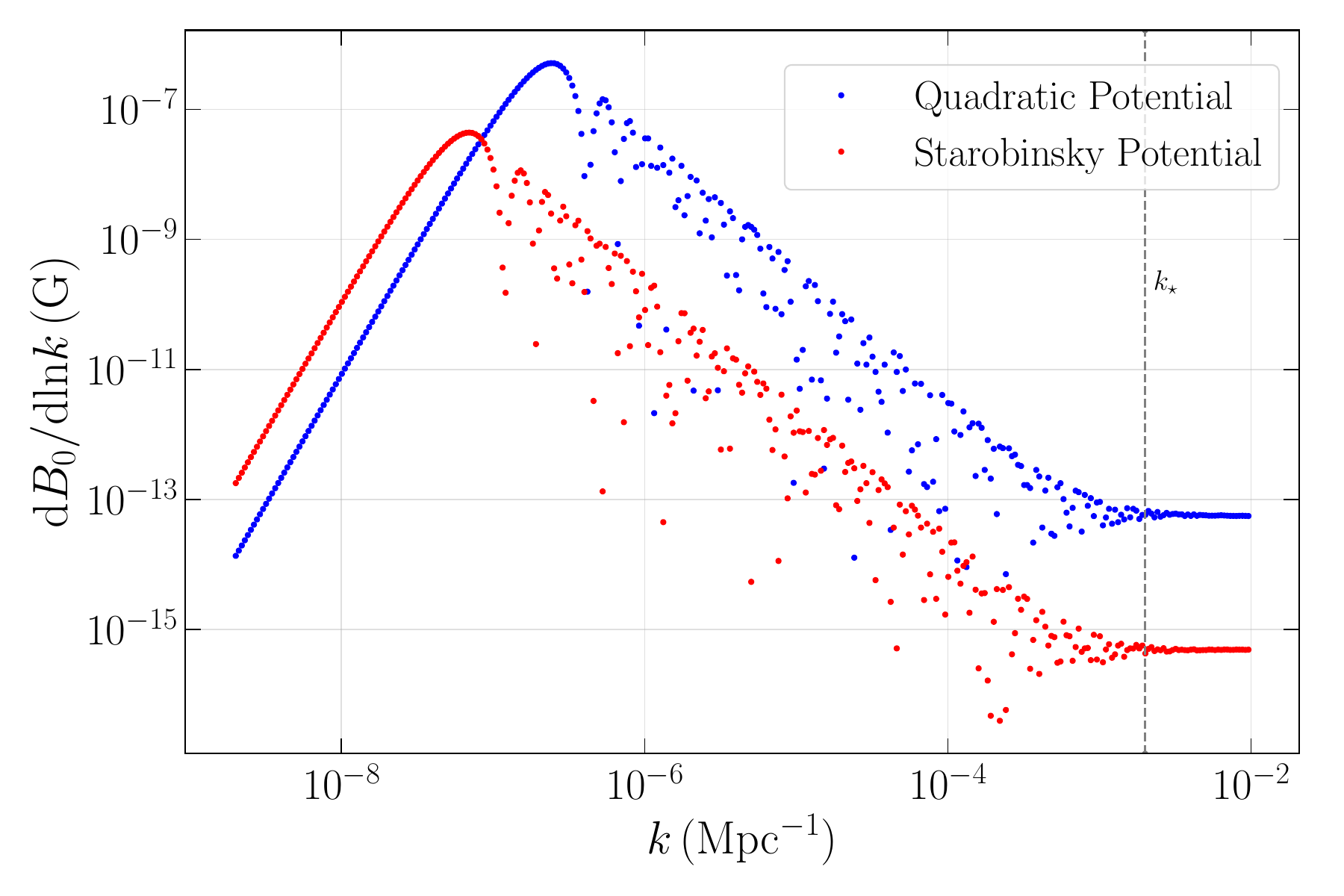}
\caption{\label{fig:B0}
Figure plots the magnetic field observable today per logarithmic interval of wave numbers, as defined in Eqn. (\ref{eqn:dB0}), for both quadratic and Starobinsky models. We have considered the coupling function of the form Eqn. (\ref{eqn:fa}) and imposed Minkowski initial conditions at the bounce. The vertical dashed line represents the pivot scale $k_\star\,=\,0.002\, \Mpc^{-1}$. A larger value of $\phi_b$ would imply a larger amount of expansion between the bounce and the onset of inflation. This in turn will make $k_\star >> k_{LQC}$ and the features due to the bounce will be redshifted and may not be observable today. 
}
\end{figure}

 
The values quoted above are the values of the magnetic field obtained after averaging over all scales. Since the magnetic field generated in LQC is scale dependent, it is informative to look at the value of the magnetic field per logarithmic interval of wave number, defined as
\begin{equation}\label{eqn:dB0}
    \frac{{\rm d}B_0}{{\rm d} \ln k}\, =\, \frac{{\cal P}^B(k)}{\sqrt{2\,\int {\rm d} \ln k\,{\cal P}^B(k)}}\,\biggl( \frac{a_e}{a_0} \biggr)^2,
\end{equation}
where ${\cal P}^B(k)$ is the power spectrum, evaluated at the end of inflation. We plot the above expression for both models in figure \ref{fig:B0}. In obtaining this figure, we have imposed Minkowski initial condition at $t\,=0\, {\rm t}_{\rm Pl}$. This picture highlights the fact that the magnetic field generated in LQC is scale-dependent. 


\par 
We close this section by considering some cases where scale-dependent features due to the bounce may not be visible today.
As explained in section \ref{sec:3}, the amount of expansion between bounce and the time at which pivot scale leaves the horizon is determined by the value of $\phi_b$. We have worked with the minimum value of $\phi_b$ which ensures that any departure from scale invariance in primordial scalar power spectrum is only imprinted on wave numbers smaller than pivot scale, $k_\star\,=\,0.002\,\Mpc^{-1}$. Such a value of $\phi_b$ imply $k_\star \approx 3\,k_{\rm LQC}$. We see that for this value of $\phi_b$, the magnetic power spectrum is also scale independent for modes larger than $k_\star$. A larger value of $\phi_b$ will blue-shift $k_\star$ with respect to $k_{\rm LQC}$. In such a scenario, it is possible that we may not observe the highly scale-dependent part of the magnetic field, and the field we measure will be similar to that generated in slow roll. We also note that the primordial magnetic field observed today can also become consistent with that generated in slow-roll scenarios for other initial conditions, such as the one corresponding to the Minkowski initial state imposed at $t\,=\,1\tpl$.

\section{Summary and Discussion\label{sec:6}}
In this work, we investigated magnetogenesis in LQC, a scenario in which inflation is preceded by a quantum bounce. We considered the electromagnetic field to be a test field living on a background dominated by a scalar field. We worked with quadratic as well as Starobinsky potentials. 
In order to break conformal invariance, we coupled the electromagnetic field to the background through a function of the form Eqn. (\ref{eqn:fa}) and studied the evolution of the field. We computed magnetic and electric power spectra at the end of inflation, explored various forms of coupling functions, investigated backreaction, and estimated the amount of primordial magnetic field that is observable today. 
\par 
If we consider Minkowski initial states imposed at the time of the bounce, then we find that the magnetic power spectrum, see figure \ref{fig:PBPE}, is scale dependent at scales with wave numbers comparable to or smaller than the scale associated with the bounce. Modes with larger wave numbers have a nearly scale-invariant magnetic power spectrum. In section \ref{sec:5A}, we explained the shape of the power spectrum by analyzing the form of $\sqrt{|f''/f|}$ as a function of time. We found that the quantum bounce in LQC changes the shape of $\sqrt{|f''/f|}$, which in turn modifies the state of modes with wave numbers $ k\lesssim k_b$. Hence, these modes are in an excited state at the onset of inflation. These excitations make the power spectrum depart from the nearly scale-invariant behaviour found in slow-roll inflation.
We further verified this by expressing the magnetic power spectrum in terms of Bogoliubov coefficients obtained by relating Minkowski states imposed at a time $t_{\rm mink}$ after the bounce with the states obtained by imposing Minkowski initial conditions at the bounce. We also found that the electric power spectrum generated in this scenario is much smaller than the magnetic power spectrum, as in the case of inflation. 
We computed the energy density of magnetic fields in section \ref{sec:5D}. We find that, in the case of the Starobinsky model, the magnetic energy density contributes only $0.1\%$ to the total energy density. Hence, we conclude that the backreaction is negligible and we can consider the electromagnetic field as a test field in this scenario. On the contrary, in the case of the quadratic potential, we find that magnetic energy density contributes about $14\%$ to the total energy density and hence is not insignificant. In section \ref{sec:5C}, we also studied equivalent forms of coupling functions. We find that, in LQC, the coupling functions of the form (\ref{eqn:fphiqp}) and (\ref{eqn:fphisp}), which are equivalent to Eqn. (\ref{eqn:fa}) during slow-roll, leads to larger magnetic fields. However, the coupling functions written in terms of Ricci scalar Eqn. (\ref{eqn:fRtrivial}) and scalar field Eqn. (\ref{eqn:fphitrivial}) are equivalent to Eqn. (\ref{eqn:fa}) even in the context of LQC. We would like to reiterate that all the above results were obtained upon imposing Minkowski initial states at the time of bounce. 
\par 
We also studied the effect of imposing Minkowski initial states at various times around the bounce. Note that this is equivalent to imposing other excited states at the bounce. The magnetic power spectra generated with such initial states for quadratic and Starobinsky potentials are plotted in figures \ref{fig:PBITQP} and \ref{fig:PBITSM}, respectively. We find that the power spectra depend on the time at which initial states are imposed, or equivalently, on the choice of initial states imposed at the bounce. For certain initial states, especially those obtained by imposing Minkowski initial conditions before the bounce, the power spectrum and hence the energy density are larger. This is particularly evident in the case of the Starobinsky model. We find that such initial states can lead to significant backreaction. On the other hand, the power spectra and hence the energy density of modes obtained by imposing the Minkowski initial conditions after the bounce are smaller. 
\par 
Finally, in section \ref{sec:5E}, we computed the residual primordial magnetic field observable today. We find that if Minkowski initial states are imposed at the bounce, observable magnetic field, $B_0\,=\, 0.94 \mu{\rm G}$ and $B_0\,=\, 0.08 \mu{\rm G}$ respectively for quadratic and Starobinsky potentials. We also note that if we consider the case where Minkowski initial condition is imposed at $t\,=\,1\tpl$, the magnetic field observable today becomes of the order of nano gauss. Planck has constrained the magnetic field observable today to be less than a few nano gauss \cite{Planck:2015zrl}. We note that the magnetic field obtained upon imposing the Minkowski initial condition at $t\,=\,1\tpl$ is consistent with this constraint. Moreover, we also noted that the observability of the scale-dependent part of the magnetic field depends on the value of $\phi_b$. For a larger value of $\phi_b$, the features arising due to the bounce will get redshifted and hence may not be observable. In such a case, the magnetic field obtained by imposing the Minkowski initial condition at the bounce may also become viable. 
\par 
As mentioned in the introduction, there are at least two reasons for studying magnetogenesis in the context of LQC. To understand if any or what signatures of the quantum bounce will be carried by the magnetic fields, and to find out whether magnetic fields generated in LQC are consistent with the current constraints on primordial magnetic fields. Our analysis shows that the primary effect of the quantum bounce in LQC is to introduce a strong scale dependence in the primordial magnetic field as opposed to the nearly scale-invariant magnetic field generated in slow-roll inflation. 
The quantum bounce modifies the initial states of the modes before they leave the horizon during inflation. These excitations manifest as scale dependence in the magnetic power spectrum. 
Our analysis also reveals that, despite the potential, there are initial states that can lead to scale-dependent magnetic fields, which are in agreement with the observable constraints. 
This work, thus, opens an intriguing possibility, namely, the ability to investigate 
quantum bounce in LQC using primordial magnetic fields. 
There are several avenues one could explore. 
\par 
For instance, there have been proposals for initial conditions based on fundamental principles for primordial scalar perturbations at or around the bounce \cite{Ashtekar:2016pqn, Ashtekar:2016wpi, Ashtekar:2020gec, Ashtekar:2021izi, Martin-Benito:2023nky}. It would be interesting to extend these proposals to the vector potential and investigate the nature of the magnetic field generated with these initial conditions in LQC. Such investigations can help us arrive at the principles that could lead us to a well-motivated initial state. 
Second, it is known that anisotropies in the background spacetime grow as the universe contracts. Hence, it is possible that spacetime at the time of the bounce is anisotropic. Since the inflationary expansion that follows the bounce dilutes this anisotropy, today's background spacetime will be isotropic. However, the primordial perturbations retain the imprints of this anisotropic phase \cite{Agullo:2020iqv, Agullo:2020wur}.  It would be interesting to explore the effects of anisotropic quantum bounce on primordial magnetic fields. Moreover, the presence of anisotropy during the quantum regime breaks the conformal invariance of the electromagnetic action. This may provide an alternate mechanism for magnetogenesis in LQC. 
Moreover, Planck had assumed a power-law form of the magnetic power spectrum in obtaining constraints on magnetic fields, whereas the scale dependence of the magnetic power spectrum generated in LQC cannot be described by a simple power law. 
Analysing imprints of such a spectrum could help us in arriving at more realistic constraints. 

\section*{Acknowledgement}

We thank Ivan Agullo and L. Sriramkumar for their comments.


\bibliography{Refs}

\end{document}